\documentclass[]{article}

\usepackage[dvipdfmx]{graphicx}
\usepackage{authblk}
\usepackage{balance}
\usepackage{url}
\usepackage{bm}

\title{How could we ignore the lens and \\pupils of eyeballs: Metamaterial optics \\for retinal projection}

\author[1, 2]{Yoichi Ochiai\thanks{wizard@slis.tsukuba.ac.jp}}
\affil[1]{University of Tsukuba}
\affil[2]{Pixie Dust Technologies, Inc.}

\date{}

\begin{document}

\maketitle

\section{Introduction: \\Democratization of Retinal Projection}

Retinal projection is required for xR applications that can deliver immersive visual experience throughout the day. If general-purpose retinal projection methods can be realized at a low cost, not only could the image be displayed on the retina using less energy, but there is also the possibility of cutting off the weight of projection unit itself from the AR goggles. Several retinal projection methods have been previously proposed; however, as the lenses and iris of the eyeball are in front of the retina, which is a limitation of the eyeball, the proposal of retinal projection is generally fraught with narrow viewing angles and small eyebox problems. In this short technical report, we introduce ideas and samples of an optical system for solving the common problems of retinal projection by using the metamaterial mirror (plane symmetric transfer optical system). Using this projection method, the designing of retinal projection can becomes easy, and if appropriate optics are available, it would be possible to construct an optical system that allows the quick follow-up of retinal projection hardware.

\newpage

\begin{figure}[t]
  \centering
  \includegraphics[width=\linewidth]{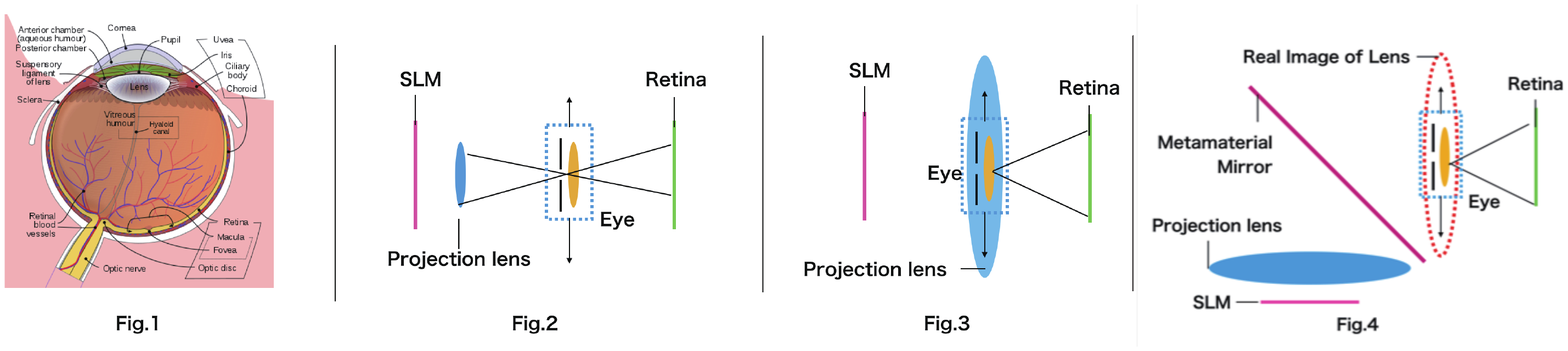}
\end{figure}

\section{Problem: \\Movable Varifocal Lens and Movable Auto \\Aperture in front of Retinal Screen}

Several methods for projecting onto the retina without the influence of the lenses and iris of the eyeballs exist. Some such methods include a projection system~\cite{Kollin:1993} that passes through the center of the lens, a projection system that uses the lens of the eye for Fourier transformation~\cite{Shi:2017}, and an imaging system that includes the focus adjustment of the eye lens~\cite{Viirre:1998}. There are various studies in this field; however, all of them have common problems. Their objective is to project an image on the retinal screen, but there exist ``variable iris'' and ``variable focus lenses'' that move in the x and y directions. Fig.1 shows the image of an eye, and Fig.2 shows the equivalent optical system of an eye. It is easy to understand that an equivalent circuit is an imaging system with a lens in the optical path moving along the x and y directions and changing its focus along the z-axis. Owing to this lens, the projected image is distorted, blurred, or masked. Furthermore, the full image is not projected on the retina when the iris moves or the hole in the light path becomes narrow.
We thus propose the use of the optical system shown in Fig.3. With the placement shown in this figure, the image does not change even if the lens moves, and the field of view is not lost even if the iris moves. However, it is difficult to place an eyeball itself in a material projection lens. However, this might be possible with real lenses using metamaterial mirrors ~\cite{Otao:2018, Otao:2017}. In the previous method, we observe the image by forming an aerial lens and looking into it. Instead, if a retina projection system that places the aerial real image lens of the projector at the eyeball position is set up, the use of a physical lens becomes unnecessary. Even if the iris moves, the viewing angle of the image does not change although the amount of light changes. In addition, the effect of the variable focus lens can be reduced. The size of the eyebox is the same as the lens size at the end of the projection system. In addition, as the ghost image that caused problems in [4] and [5] would be cut by the iris, there is a possibility that a clearer image could be formed on using this method. Fig.4 shows the optical schematics of the aerial image and eye.


\section{Experimental Result and Discussion}

As shown in the optics setup in Fig.5, even if an aperture or lens is placed at the aerial projection lens, the projection image is clearly visible on the screen even if the aperture or lens is moved. The image captured with a camera at the aerial image of the projection lens is shown in Fig.6. The ghost image is cut by the aperture. This result indicates the possibilities of realizing a transmissive optical system for retinal projection with a large eyebox. The aerial optical system at the ``focal point'' of the TMD is essentially very similar to the pinlight display [6]; however, the pinlight that forms the aerial image of the projection lens is arranged as the number of aerial light source as the same number of grids of the metamaterial mirror. Therefore, the viewing angle is subject to constraints on the incident angle and reflection angle of the metamaterial mirror. Although, theoretically, the viewing angle is limited to approximately 90 degrees by a single planner mirror. In contrast, it is not necessary to physically arrange the array of pinlights in front of the eyes, and the density of the light source can be large. Moreover, it can be placed in the eyeball. Therefore, it is difficult for the image to be affected by the eye lens and iris. In addition to improving the design possibilities, the calculation cost is also reduced. This method is just a 2D image projection that ignores the effect of the lens and iris. There are several issues that are yet to be addressed such as the subject experiments using this method and the presentation of detailed reports on the implementation systems, and the methods for reducing its size in the production process. We hope that these subjects will be reported on in the near future.

\begin{figure}[t]
  \centering
  \includegraphics[width=\linewidth]{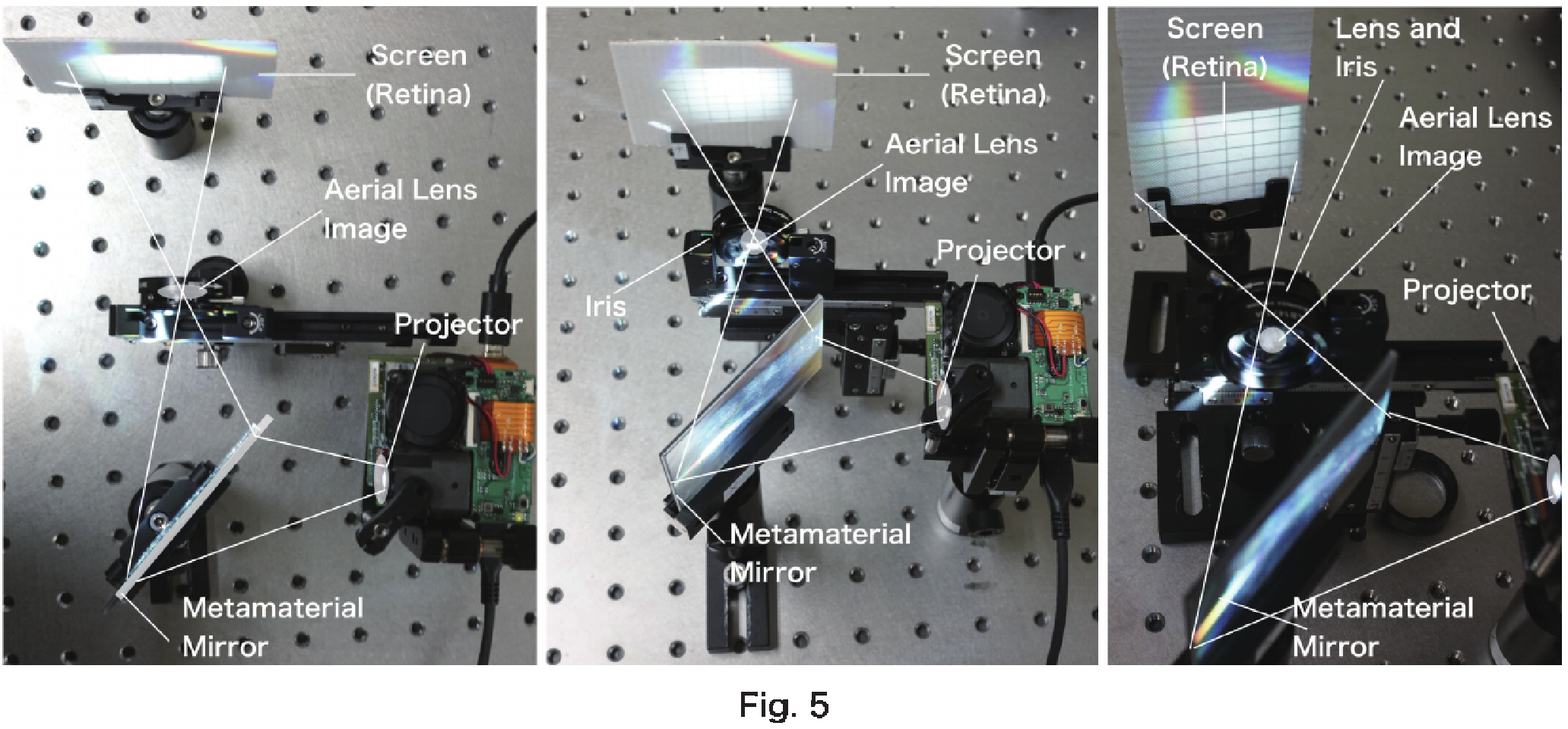}
\end{figure}

\begin{figure}[t]
  \centering
  \includegraphics[width=\linewidth]{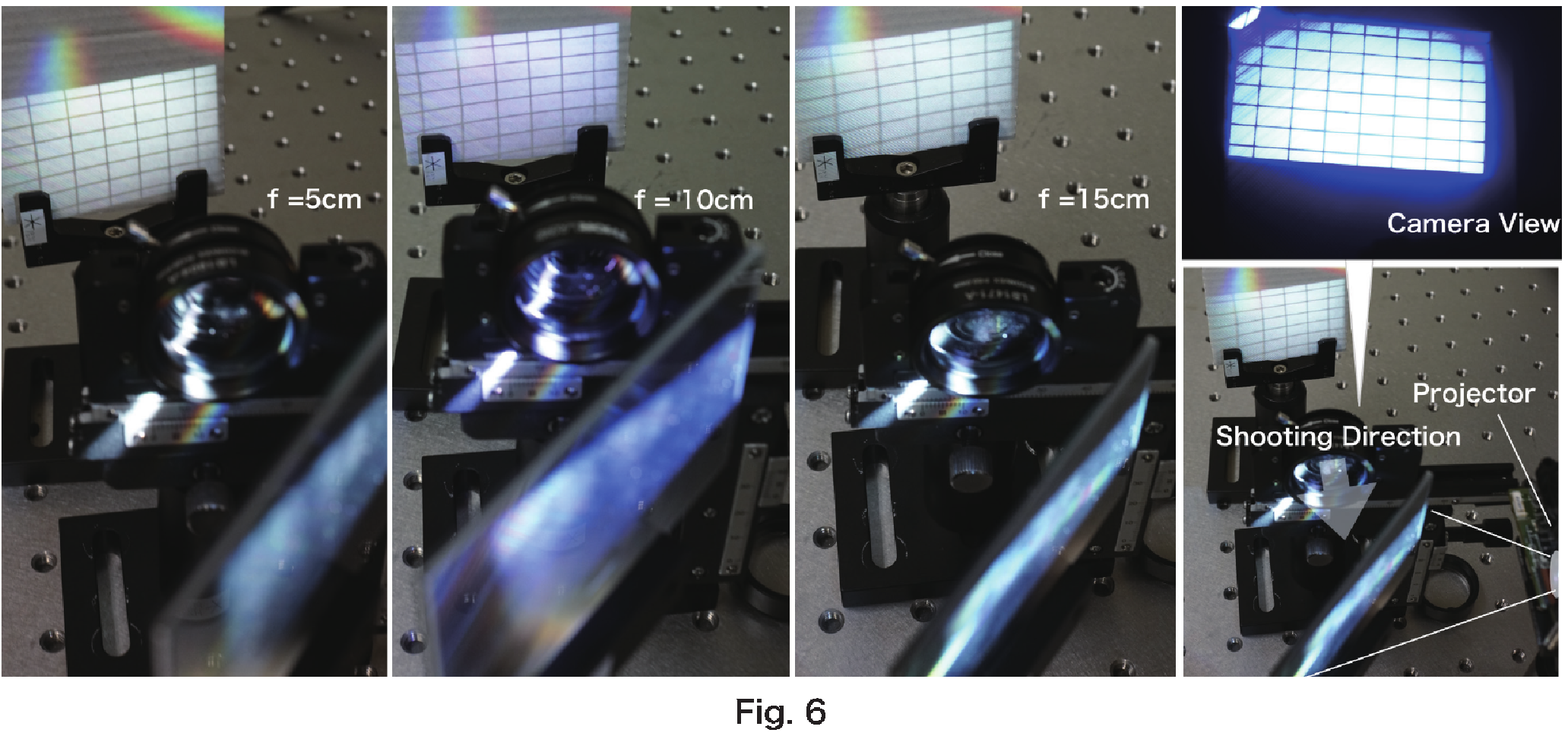}
\end{figure}

\bibliographystyle{unsrt}
\bibliography{2018-arxiv-retinal}

\end{document}